
\input harvmac.tex
\noblackbox

\def\s{\sigma}

\lref\conclass{P. Green and T. Hubsch, ``Phase Transitions Among
(Many of) Calabi-Yau Compactifications'', {\it Phys. Rev. Lett.} 
{\bf 61} (1988) 1163; P. Candelas, P. Green, and T. Hubsch, ``Rolling
Among Calabi-Yau Vacua'', {\it Nucl. Phys.} {\bf B330} (1989) 49.}
\lref\strom{A. Strominger, ``Massless Black Holes and Conifolds
in String Theory'',{\it Nucl. Phys.} {\bf B451} (1995) 96,
hep-th/9504090}
\lref\sminst{E. Witten, ``Small Instantons in String Theory'',
{\it Nucl. Phys.}
{\bf B460} (1996) 541, hep-th/9511030}
\lref\otherNone{C. Vafa and E. Witten, ``Dual String Pairs with
N=1 and N=2 Supersymmetry in Four Dimensions'', 
hep-th/9507050; J. Harvey, D. Lowe,
and A. Strominger, ``N=1 String Duality'' 
{\it Phys. Lett.} {\bf B362} (1995) 65, hep-th/9507168; 
C. Vafa and A. Sen, ``Dual Pairs of Type II String
Compactification'', {\it Nucl. Phys.}
{\bf B455} (1995) 165, hep-th/9508064}
\lref\gaugcond{S. Kachru and E. Silverstein,
``N=1 Dual String Pairs and Gaugino Condensation'', hep-th/9511228; to
appear in {\it Nucl. Phys.} {\bf B}.}
\lref\phases{E. Witten, ``Phases of N=2 Theories in Two Dimensions'',
{\it Nucl. Phys.} {\bf B403} (1993) 159, hep-th/9301042}
\lref\dkI{J. Distler and S. Kachru, ``(0,2) Landau-Ginzburg Theory'',
{\it Nucl. Phys.} {\bf B413} (1994) 213, hep-th/9309110. }
\lref\dktwo{J. Distler and S. Kachru, ``Singlet Couplings and 
(0,2) Models,'' {\it Nucl. Phys.} {\bf B430} (1994) 13, hep-th/9406090.}
\lref\SilvWit{E. Silverstein and E. Witten, ``Criteria for Conformal
Invariance of (0,2) Models'', {\it Nucl. Phys.} {\bf B444} (1995) 161,
hep-th/9503212}
\lref\BCOV{M. Bershadsky, S. Cecotti, H. Ooguri, and C. Vafa,
``Kodaira-Spencer Theory of Gravity and Exact Results for Quantum
String Amplitudes'', {\it Comm. Math. Phys.} {\bf 165} (1994) 311.}
\lref\KL{V. Kaplunovsky and J. Louis, ``On Gauge Couplings
in String Theory'', {\it Nucl. Phys.} {\bf B444}
(1995) 191.}
\lref\BSV{M. Bershadsky, V. Sadov, and C. Vafa, ``D-Branes
and Topological Field Theory'', hep-th/9511222.}
\lref\KMP{S. Katz, D. Morrison, and R. Plesser, ``Enhanced Gauge
Symmetry in Type II String Theory'', hep-th/9601108.}
\lref\Seimod{N. Seiberg, ``Exact Results on the Space
of Vacua of Four-Dimensional SUSY Gauge Theories'',
{\it Phys. Rev.} {\bf D49} (1994) 6857, hep-th/9402044.}
\lref\ntwodual{S. Kachru and C. Vafa, ``Exact Results for
N=2 Compactifications of Heterotic Strings,'' {\it Nucl. Phys. }
{\bf B450} (1995) 69, hep-th/9505105; S. Ferrara, J. Harvey, A.
Strominger, and C. Vafa, ``Second-Quantized Mirror Symmetry,'' 
Phys. Lett. {\bf B361} (1995) 59, hep-th/9505162}
\lref\CDGP{P. Candelas, X. de la Ossa, P. Green, and L. Parkes,
``A Pair of Calabi-Yau Manifolds as an Exactly Solvable
Superconformal Field Theory'', {\it Nucl. Phys.} {\bf B359}
(1991) 21.}
\lref\ADS{I. Affleck, M. Dine, and N. Seiberg,``Dynamical
Supersymmetry Breaking in Supersymmetric QCD'',{\it Nucl. Phys.}
{\bf B241} (1984) 493. } 
\lref\GP{B. Greene and R. Plesser, ``Duality in Calabi-Yau
Moduli Space'',{\it Nucl. Phys.}
{\bf B338} (1990) 15.}
\lref\MP{D. Morrison and R. Plesser, ``Summing the Instantons:
Quantum Cohomology and Mirror Symmetry in Toric Varieties'',
{\it Nucl. Phys.}
{\bf B440} (1995) 279, hep-th/9412236}
\lref\split{C. Okonek, M. Schneider, and H. Spindler, 
{\it Vector Bundles on Complex Projective Spaces}, (Birkhauser, 1980).}
\lref\Katz{S. Katz, ``On the Finiteness of Rational Curves on 
Quintic Threefolds'', {\it Comp. Math.} {\bf 60} (1986) 151.}
\lref\DMW{M. Duff, R. Minasian, and E. Witten, ``Evidence
for Heterotic/Heterotic Duality'', hep-th/9601036.}

\Title{\vbox{\hbox{HUTP-96/A016}\hbox{PUPT-1622}\hbox{RU-96-30}
\hbox{\tt hep-th/9605036}}} 
{\vbox{\centerline{SUSY Gauge Dynamics and}
\medskip
\centerline{Singularities of $4d$ N=1 String Vacua}}}
\centerline{Shamit Kachru\footnote{$^*$}
{kachru@string.harvard.edu}}
\smallskip\centerline{\it Lyman Laboratory of Physics, Harvard
University, Cambridge, MA 02138}
\smallskip
\centerline{Nathan Seiberg\footnote{$^{**}$}
{seiberg@physics.rutgers.edu}}
\smallskip\centerline{\it Department of Astronomy and Physics,
Rutgers University, Piscataway, NJ, 08855}
\smallskip
\centerline{Eva Silverstein\footnote{$^{***}$}
{silver@puhep1.princeton.edu}}
\smallskip\centerline{\it Joseph Henry Laboratories, Princeton University,
Princeton, NJ 08544}
\vskip .2in

Many N=1 heterotic string compactifications exhibit physically
mysterious singularities at codimension one in the moduli space of
vacua.  At these singularities, Yukawa couplings of charged fields
develop poles as a function of the moduli.  We explain these conformal
field theory singularities, in a class of examples, as arising {}from
non-perturbative gauge dynamics of non-perturbative gauge bosons (whose
gauge coupling is the sigma model coupling) in the string theory.

\Date{May 1996} 

\newsec{Introduction}

Recent progress in string duality raises hope for a quantitative
nonperturbative understanding of generic string vacua.  Much progress
has been made in the special classes of models with extended
supersymmetry (in the $4d$ sense).  In particular, previously mysterious
singularities in conformal field theory moduli space (such as conifold
singularities arising in type II theories compactified on Calabi-Yau
manifolds \strom\ and small instanton singularities in $SO(32)$
heterotic and type I theories \sminst) have been explained as the result
of additional (sometimes solitonic) states becoming massless at the
singularity.  For models with $4d$ N=1 supersymmetry much less is known,
as the supersymmetry is much less constraining.  Some examples of dual
pairs exist \refs{\otherNone,\gaugcond}, but these examples are rather
special 
as they arise as orbifolds of dual pairs with extended supersymmetry.
It is the purpose of the present paper to begin the study of more
generic $N=1$ string compactifications, by addressing the issue of
singularities in the moduli space for a class of these models.

A heterotic $N=1$ model is obtained by compactification on an internal
(0,2) superconformal field theory.  Such models have been fruitfully
studied using a gauged linear sigma model on the worldsheet
\refs{\phases, \dkI},
which flows in the infrared to a (0,2) superconformal field theory
\SilvWit.  Singularities in the conformal field theory arise at
codimension one loci in the linear sigma model moduli space where some
multiplet(s) become free on the worldsheet, leading to a divergence in
the path integral \phases.  This leads in particular to a simple pole in
a Yukawa coupling $\kappa$ \SilvWit.  For the special case of (2,2)
theories, this is the familiar pole which occurs at the conifold
\CDGP.  If $z$ is a coordinate on the moduli space such that 
$z\to 0$ is the singularity, then we find
\eqn\yukpole{\kappa\sim{{g^3}\over{z}}f({\rm Moduli})}
where $g^3$ is the invariant coupling of the relevant charged fields and
$f$ is some holomorphic function of the moduli, nonsingular in the
limit.

Because the conformal field theory has become singular,
it is natural to suspect that extra states are becoming light.
Their dynamics should then explain the singular behavior of
the conformal theory. 
The singularity \yukpole\ of the classical $N=1$ theory cannot
arise {}from a perturbative effect, as occurred in the 
resolution of the analogous logarithmic singularity
at the $N=2$ conifold \strom.  Instead, the quantum resolution here
must involve nonperturbative dynamics of the new light particles.

Poles in the superpotential 
do arise dynamically in supersymmetric
QCD with the number of flavors one less than the number of colors \ADS.      
In this paper we will discover a detailed relation between
these two sets of poles.  In section 2 we explain how a simple
asymptotically free enhancement to the spectrum, $SU(2)$ with four
doublet chiral multiplets, can reproduce the singularity \yukpole,
with a particular function $f$.  In section 3 we discuss
in general terms a class of N=1 heterotic string models--
compactifications of the $SO(32)$ heterotic string theory on Calabi-Yau 
threefolds which are $K3$ fibrations--with codimension
one ``conifold'' singularities.  
We derive the spectrum of light particles at the singularity and 
find it agrees with that discussed in \S2!  

The derivation of the spectrum in such a model proceeds as follows.  On
the (0,2) moduli space, there are codimension one singularities which
consist of a small instanton on the K3 fiber fibered over the ${\bf
P^1}$ base C.  Because the nonperturbative enhancement of the spectrum
of the $6d$ theory compactified on the generic fiber is known \sminst,
we can deduce the spectrum for the $4d$ $N=1$ theory on the fibration
using similar methods to those employed in the $N=2$ case in \refs{\BSV,
\KMP}.  That is, we compute the massless spectrum in four dimensions by
determining the zero modes of the relevant Dirac operator on C.

The result one obtains for this class of models is a nonperturbative
enhancement of the spectrum consisting of $SU(2)$ gauge symmetry with
four doublets.\foot{This result applies on the generic (0,2) locus,
where the six-dimensional singularity is a small instanton.  It
would be interesting to determine the behavior on the (2,2) locus,
where for example the tree-level metric takes the $N=2$ form.}  
Instanton effects in the $SU(2)$ gauge theory explain
the singularities of the conformal field theory.  Note that the
singularity occurs at string tree level in the heterotic theory.  This
can be reproduced by the dynamics of the $SU(2)$ gauge bosons since they
are non-perturbative, and their interactions are governed by the sigma
model coupling.  When the type I theory is compactified on the same
space\foot{In a compactification to six dimensions on K3 this theory is
dual to the corresponding heterotic theory.  However, in four dimensions
this might not be the case.  The generic theory of this kind has
``anomalous $U(1)$'' factors and develops Fayet-Iliopoulos D-terms
\ref\dsw{M. Dine, N. Seiberg and E. Witten,
``Fayet-Iliopoulos Terms in String Theory,'' {\it Nucl. Phys.} {\bf
B289} (1987) 589.}
which lead to a shift in the vacuum.  It is not clear to us whether
heterotic-type I duality is true in this case.}, the
extra states arise perturbatively in the DD and DN sector of open
strings in a fivebrane background, and the dynamically induced
superpotential is nonperturbative in $g_{string}$.  The situation is
very analogous to the one encountered in 4d N=2 duality where the type
II sigma model sums up nonperturbative effects for the heterotic string
vector multiplets \ntwodual.  
Here, the heterotic string sigma model computes nonperturbative 
corrections which can be ascribed to perturbative type I states. 
In section 4 we present in some detail an explicit example.

\newsec{SUSY QCD and the Singularity}

Supersymmetric QCD with $N_F=N_C=2$ has a smooth quantum moduli
space of vacua which is a deformation of the classical one.
If we denote the four doublets $d_i$, $i=1,\dots,4$, we can
form gauge-invariant coordinates 
$V_{ij}=d_i^\alpha\epsilon_{\alpha\beta}d_j^\beta$ where
$\alpha,\beta$ are $SU(2)$ indices.  Then the classical
moduli space is given by 
$Pf(V)=\epsilon^{ijkl}V_{ij}V_{kl}=0$.
Quantum mechanically, a one instanton effect removes   
the singularity:
\eqn\qmod{Pf(V)=\Lambda^4}
where $\Lambda$ is the dynamical scale of the gauge theory \Seimod.
The moduli space is five-complex-dimensional (with 
the $SU(2)$ gauge symmetry Higgsed at generic points).
The constraint \qmod\ can be enforced with a Lagrange multiplier
term in the superpotential:  
\eqn\supone{W_0=\lambda(Pf(V)-\Lambda^4)}

In our problem there is one relevant dimension in the moduli space,
the coordinate $z$ \yukpole.  For this to agree with the gauge theory
we need to add certain tree-level nonrenormalizable interactions 
to the superpotential \supone.  In particular, consider adding
the terms $V_{13}^2+V_{23}^2+V_{14}^2+V_{24}^2$ (the particular form of
these terms is not crucial for our analysis).  Integrating
out $V_{13},V_{23},V_{14}$, and $V_{24}$ we find that they
all vanish.  Then the Pfaffian constraint obtained by integrating
out $\lambda$ requires $V_{34}={\Lambda^4\over V_{12}}$,
yielding  
a one-dimensional moduli space.

Now what about the pole?  Consider adding the term $g^3V_{34}$ to
the superpotential so that altogether we have

\eqn\supfull{\eqalign{
W=
& g^3V_{34}+\lambda(V_{12}V_{34}-V_{13}V_{24}+V_{14}V_{23}
-\Lambda^4)\cr
&+V_{13}^2+V_{23}^2+V_{14}^2+V_{24}^2\cr}}  
Integrating out $V_{13},V_{23},V_{14}$, and $V_{24}$ we still
find that they all vanish.  Furthermore, the constraint 
$V_{34}={\Lambda^4\over V_{12}}$  yields upon substitution
\eqn\supfin{W_{dyn}={g^3\Lambda^4\over{V_{12}}}}
This is the desired pole, where the function $f$ {}from
\yukpole\ is given by $\Lambda^4$ as a function of moduli.  This
will be elucidated in the following sections, where we will 
see this phenomenon realized in $K3$ fibration 
models.\foot{This is assuming that tree-level nonrenormalizable
interactions of the sort in \supfull\ arise given the spectrum;
we have not performed the necessary computations to check this
directly, but such terms must arise in order to reproduce the
correct dimension of the moduli space.}

Because it occurs for $V_{34}\to\infty$, in the gauge theory this pole
appears to be at infinite distance in the moduli space.  This would be
in contradiction with known results for (2,2) models \conclass.  In our
problem this gauge theory occurs as a nonperturbative enhancement at the
singularity in the conformal field theory.  The validity of the gauge
theory analysis is limited to vacuum expectation values $V_{ij}<<M_S^2$
where $M_S$ is the string scale.  Therefore, the distance to the pole is
not determined by this analysis.  To be precise, this limitation is
determined as follows.  At large $V_{34}$, the Kahler potential for the
theory
\supone\ is given by its classical value (since all the
fields are Higgsed):  $K\sim |V_{34}|$.  
The gauge theory breaks down for $V_{34}\sim M_S^2$ or
equivalently $V_{12}\sim{\Lambda^4\over{M_S^2}}$.  
The distance {}from this point to a finite point $A$ in the moduli space
is 
\eqn\dist{d\sim\int_{\Lambda^4\over{M_S^2}}^A{{d|V_{12}|}\over
{|V_{12}|^{3\over 2}}}\Lambda^2=-{\Lambda^2\over{\sqrt{A}}}+M_S}
which is of course finite.
The distance to the pole is not calculable in the field theory
approximation. 

\newsec{Singularities in K3 Fibrations and SUSY QCD}

In this section we will discuss in general terms the string models for
which the desired spectrum at the singularity, $SU(2)$ with four
doublets, can be derived.  In the next section we will provide the
details for an example.  Consider the $SO(32)$ heterotic or type I
string compactified on a K3 fibration, with holomorphic vector bundle V.
In the adiabatic limit of large ${\bf P^1}$ base size, such a model
appears locally like a compactification to six dimensions on the fiber
theory.  Singularities can occur when the vector bundle or manifold
becomes singular.  In general one might expect singularities at
codimension one to occur at isolated points in the manifold.

In fact, for
K3 fibrations there is a codimension one component of the singular 
locus for which the theory on the generic fiber is singular.
On the (2,2) locus, for example, one finds at codimension one
an $A_1$ singularity fibered over the base, as studied by \refs{\BSV,
\KMP} in compactifications of type II string theory.  
In the (0,2) context, singularities generically involve the
vector bundle but not the manifold becoming singular.  The
codimension one singularity of the generic fiber in a (0,2) model
will consist of a single small instanton.  

It was demonstrated in \sminst\ that the six-dimensional
theory obtained by compactifying the heterotic or type I string
on K3 with such a singularity in the vector bundle has
an enhanced symmetry:
One finds SU(2) with hypermultiplets in the
$({\bf 32,2})$ of $SO(32)\times SU(2)$.
In the type I theory the small instanton is a D-5-brane; the
SU(2) gauge bosons come {}from the DD (or 55) sector while the
hypermultiplets arise in the DN (or 59) sector.  On the
heterotic side, the enhanced $SU(2)$ is non-perturbative
instead of arising in the perturbative string spectrum. 
In the next subsections we will show that the $SU(2)$ gauge
symmetry along with four doublets 
survives reduction to the 4d N=1 theory obtained by
fibration over a sphere C.  We will focus on
the heterotic string, and will denote this nonperturbative
enhancement of the gauge field spectrum $SU(2)_{NP}$.

Before embarking on the derivation of this spectrum, let us
discuss its relation to the gauge dynamics explained in section 2.
In six dimensions, the $SU(2)_{NP}$ gauge fields have a kinetic
term which (in the string frame) is independent of the string 
coupling \DMW.
Upon reduction on C with radius $R$, this leads to an effective
four-dimensional $SU(2)$ gauge coupling ${1\over{g_{NP}^2}}\propto 
{R^2\over\alpha^\prime}$.
Then 
\eqn\lambd{\Lambda\propto M_Se^{-{R^2\over\alpha^\prime}}}   
Since the relevant part of the moduli space is one dimensional
(the single modulus given by $z$ in \yukpole ), the four
doublets must have interactions giving a tree-level superpotential
such as \supfull\ which ensures that.  It would be interesting
(though difficult) to compute these interactions in the type I
theory using conformal field theory.  Assuming such terms are there,
we obtain the result \supfin.
In particular, the function $f$ in \yukpole, which was
determined in \supfin\ to be $\Lambda^4$, depends only
on the moduli and not on the string coupling, as required for
a conformal field theory effect.  So we see that the spacetime
instanton effect \supfin\ computes for us nonperturbative
effects on the worldsheet of the heterotic string.  

Putting together the gauge theory and string theory, we have the
following hierarchy of scales.  Above the scale $1/R$ the theory is
six-dimensional and was studied in \sminst.  Below the scale $1/R$, but
above the scale $\Lambda$, the model is four-dimensional and lives on
the classical moduli space of $SU(2)$ with four doublets.  Including the
instanton effect leads to the deformation \supone\ of the moduli space,
removing the singularity at the origin.  {}From the worldsheet point of
view this reflects the fact that worldsheet instantons, which wrap
around the entire base, probe beyond the local adiabatic regime and can
alter the structure of the moduli space inherited {}from six dimensions.
Note that in the present context of $K3$ fibrations, the coupling
\yukpole\ goes to zero in the limit of $\Lambda\to 0$.  The
six-dimensional theory has no such coupling, so this form of $f$ is
consistent.

Let us now proceed with the computation of the spectrum.  As noted in
the type II context by \refs{\BSV, \KMP}, the nontrivial fibration of
the K3 theory over the base C is crucial in obtaining a consistent
supersymmetric theory in 4d, as C alone is not flat.  For $N=1$
supersymmetry we require one covariantly constant spinor to survive the
reduction. This is possible due to the twisting of the Dirac operator on
C arising {}from the Lorentz and gauge transformation properties of the
fields in the full Calabi-Yau.  The twisting to obtain $N=1$
supersymmetry occurs as follows.  The Lorentz group $SO(4)$ in the
internal four dimensions of the K3 decomposes as $SO(4)=SU(2)_H\times
SU(2)^\prime$ where $SU(2)_H$ is the holonomy group of the K3.  The
components of the six-dimensional supercharges have charges $\pm 1/2$
under the Lorentz $U(1)$ on C and transform as a ${\bf 2}$ of
$SU(2)^\prime$.  Therefore a twist by the $U(1)$ generator $J^\prime_3$
in $SU(2)^\prime$ preserves half of the $6d$ $N=1$ supersymmetry, giving
$N=1$ supersymmetry in four dimensions.  Meanwhile the $N=1$ SU(2)
vector multiplets are singlets under Lorentz and $SO(32)$ gauge
symmetries, as well as under the Lorentz rotations on C.  Thus they
survive on reduction as constants on C.

\subsec{The Charged Matter Spectrum {}from the Splitting Type of V}

More analysis is required to find the components of the 6d
hypermultiplets which survive as chiral multiplets in four dimensions.
In the type I theory, the nonperturbative spectrum lives on a D-brane
which is partially wrapped around the curve C.  The origin of the
hypermultiplets in quantizing the D-brane is explained in section 3.1 of
\sminst.  Let 2,...,5 be the coordinates on the 5-brane in light cone
gauge and 6,...,9 the normal coordinates.  In the DN Neveu-Schwarz
sector (which gives spacetime bosons) the vacuum is a spinor of
$SO(4)_{6,...,9}$ and a scalar of $SO(4)_{2,...,5}$.  In the DN Ramond
sector (which gives spacetime fermions) the vacuum is a scalar of
$SO(4)_{6,...,9}$ and a spinor of $SO(4)_{2,...,5}$.

Let us count the fermions; the bosons are then given by the 
surviving $N=1$ supersymmetry just discussed.
We have 32 hypermultiplets, each of which has two chiral multiplets.
Six of the 32 hypermultiplets lie in the vector bundle $V$, $V^*$
of the CY theory.  Let us first consider these: We need zero modes of
the Dirac operator acting on these fermions, which are given by
$H^0({\cal O}(-1)\otimes V)$ and
$H^0({\cal O}(-1)\otimes V^*)$.  Any holomorphic vector bundle
on ${\bf P}^1$ splits into a sum of holomorphic line bundles \split.
These line bundles are denoted ${\cal O}(k)$, where $k$ is the
integrated first Chern class $\int_{\bf P^1} c_1({\cal O}(k))=k$.
For $k\ge 0$, ${\cal O}(k)$ has $k+1$ sections; for
$k<0$ there are none. 
In general, for a rank three bundle, 
\eqn\Vsplit{V\biggl|_C={\cal O}(a)\oplus {\cal O}(b)\oplus {\cal O}(c)}
and
\eqn\coVsplit{V^*\biggl|_C={\cal O}(-a)\oplus {\cal O}(-b)\oplus {\cal
O}(-c).} 
where $c_1(V)\propto a+b+c\equiv 0$.  We will show in the
next section that for (a class of) K3 fibration models, 
$a=2$, $b=0$, and $c=-2$.  
Tensoring with ${\cal O}(-1)$, we have $N_\psi$ zero modes, where 
\eqn\zero{N_\psi=2\times \biggl(h^0[{\cal O}(1)\oplus {\cal O}(-1)
\oplus {\cal O}(-3)]
+h^0[{\cal O}(-3)\oplus {\cal O}(-1)\oplus {\cal O}(1)]\biggr)}
so 
\eqn\zerocount{N_\psi=2\times ([2+0+0]+[0+0+2])=8}
Since 8 chiral fermions makes for 8 chiral multiplets, we
will have 4 $SU(2)$ doublets in the surviving spectrum.
Note that without the above twist of the spin bundle 
${\cal O}(-1)$ by the vector bundle V with nontrivial splitting type
there would be no sections,
and none of the hypermultiplets would survive.  In particular,
the other 26 hypermultiplets do not survive, so we are left with
precisely 4 doublets.

\newsec{Explicit Example}

The best-studied class of $N=1$ heterotic string vacua consists of those
models that can be realized as the infrared limit of a gauged linear
sigma model on the worldsheet \refs{\phases, \dkI, \SilvWit}.  Many such
examples have a phase which is geometrical, in that (for large values of
a parameter) the model reduces to a nonlinear sigma model on a
compactification manifold.  This manifold K must be Calabi-Yau and
equipped with a stable holomorphic vector bundle V in order to satisfy
the large-radius conditions for conformal invariance.  Conformal
invariance throughout the linear sigma model moduli space was
demonstrated in \refs{\dktwo, \SilvWit}.  The structure of singularities
in such (0,2) models is manifest in the linear sigma model description
\phases.  In this description the moduli appear as coupling constants in
the two-dimensional sigma model.

Motivated by the considerations of the previous sections, 
let us consider an example which admits a relation to six
dimensions, namely a K3 fibration.  
One example is a (0,2) model on the hypersurface of degree 8
in ${\bf W}P_{11222}^4$.\foot{In fact, the following analysis
(with obvious modifications) 
will hold for all K3 fibration models with weights of
the form $(1,1,2k_1,2k_2,2k_3)$.}  Take the defining equation
\eqn\manG{G_0(z_1,\dots, z_5)
={z_1^{8}\over 8}+{z_2^{8}\over 8}+{z_3^4\over 4}+
{z_4^4\over 4}+{z_5^4\over 4}+\dots=0.}
This hypersurface has a 
curve of $A_1$ singularities inherited {}from the ${\bf WP}_{11222}^4$:
\eqn\wpsing{\biggl\{z_i=(0,0,z_3,z_4,z_5)| {z_3^4\over 4}+
{z_4^4\over 4}+{z_5^4\over 4}+\dots=0\biggr\}}
Resolving this singularity involves inserting a ${\bf P}^1$ for each
singular point on the curve.  The precise procedure will become
clear momentarily {}from the 
construction of the linear sigma model for this compactification.
To be specific we will take V to be the rank 3 deformation of the
tangent bundle on this hypersurface.

The (0,2) linear sigma model describing compactification on
this hypersurface has the following multiplets and interactions.
We will work in $(0,2)$ superspace
with fermionic coordinates $\theta^+$ and $\bar\theta^+$ and bosonic
coordinates $y^\alpha$, $\alpha=1,2$. 
The imaginary part of the Kahler 
parameter, $r_1$, arises as the coefficient
of a Fayet-Iliopolous $D$-term of a world-sheet $U(1)$ gauge group
with Fayet-Iliopolous parameter $r_1$.
There is an additional Kahler parameter $r_2$
which determines the size of the ${\bf P}^1$ resolving \wpsing, leading
to a second U(1) gauge multiplet on the worldsheet.  We
will call these two Abelian factors $U(1)_A$, where
$A=1,2$.
 
The superpotential terms will involve seven chiral superfields.
The first set corresponds to the coordinates $z_i=(z_1,\dots,z_5)$:
\eqn\zmult{Z^i=z^i+\sqrt{2}\theta^+\psi^{i}_+
-i\theta^+\bar\theta^+(D_0+D_1)z^i,\,\, i=1,\dots 5}
These will have charges (1,1,2,2,2) under $U(1)_1$ and 
(1,1,0,0,0) under $U(1)_2$.  
Because we are considering a (0,2) deformation of
a (2,2) model, each chiral multiplet has an associated
left-moving fermionic multiplet  
\eqn\fmult{\Lambda_-^{i}=\lambda_-^i-\sqrt{2}\theta^+G^i-
i\theta^+\bar\theta^+(D_0+D_1)\lambda_-^{i}-
2i\bar\theta^+Q^A_i\Sigma^\prime_AZ^i} 
with the same
gauge charges $Q^i_A$ as the corresponding $Z^i$. 
Here $G^i$ is an auxiliary field which gets integrated out in favor
of the (0,2) superpotential term $\bar J_i$ which is introduced below.  
(In $(2,2)$ language,
the $Z^i$ and $\Lambda_-^{i}$ would combine into ordinary chiral
superfields.) 
The $\Lambda_-$ obey a chirality condition 
\eqn\hukak{\bar D_+\Lambda_-^{i} = E^i=2iQ^i_A\Sigma^\prime_A Z^i.}
where $E^i$ are holomorphic functions of the chiral superfields.

In addition we will have
two more chiral multiplets (and associated left-moving fermionic
multiplets) $(p,\lambda_-^p,\psi_+^p)$ with charges (-8, 0)
under the two U(1)'s and 
$(x,\lambda_-^x,\psi_+^x)$ with charges (0,-2).

The Lagrangian consists of the
standard flat kinetic terms together with the $U(1)$ $D$-terms
and $\theta$-terms, as well as superpotential terms.
The 
superpotential terms are 
\eqn\sup{L_J=-{1\over \sqrt{2}}\int d^2y\,\,d\theta^+
\Lambda_-^{I} J_I\biggr|_{\bar\theta^+=0} + h.c.}
where $I$ indexes the chiral superfields $Z_i,P,X$.
Here 
\eqn\defJ{J_p=G(Z^i,X)}
where $G(Z^i,X)=0$ is the defining equation for the resolved
hypersurface:
\eqn\defGfull{G(z_i,x)={z_1^8x^4\over 8}+{z_2^8x^4\over 8}+
{z_3^4\over 4}+{z_4^4\over 4}+{z_5^4\over 4}
+\dots.}
The $\dots$ refers to the other gauge-invariant terms that
can occur in the $\Lambda_-^p$ term of \sup.

In the other terms, $J_i\equiv p\tilde J_i$, 
$J_x\equiv p\tilde J_x$, where $\tilde J_i$ and 
$\tilde J_x$ are homogeneous polynomials in $z_i,x$ of the
appropriate degrees to render the terms gauge invariant.  They satisfy 
the conditions
\eqn\EcondI{Q^i_1z_iJ^i=8PG(z^i)}
and
\eqn\EcondII{z_1J^1+z_2J^2-2xJ_x=0.}
This ensures (using \hukak\ and the values of the charges)
that $\sum_I E^IJ_I=0$, so that \hukak\ is consistent with \sup\
having (0,2) supersymmetry.
The model actually has $(2,2)$ supersymmetry if and only if
$J_i=P{\partial G\over \partial Z^i}$
and
$J_x=P{\partial G\over \partial x}$
since then the superpotential term \sup\ can be written
in (2,2) superspace as $\int d^2\theta PG$.
Departing {}from this locus breaks $(2,2)$ supersymmetry 
to $(0,2)$ and has the effect of perturbing the tangent bundle
of the hypersurface to a more general bundle $V$.  
We will work explicitly with such a (0,2) deformation
below.  (Since the perturbed
bundle has rank three, the space-time gauge group is still 
$SO(26)\times U(1)$.) 
If we decompose $\tilde J_i$ as 
$\tilde J_i={\partial G\over{\partial s^i}}+G_i$ for some
homogeneous polynomials
$G_i$ with the appropriate gauge charges satisfying $G_i z^i=0$, then
the parameters in $G_i$ are the moduli that break $(2,2)$ down to
$(0,2)$.


The bosonic potential for this model is 

\eqn\bospot{\eqalign{U(\phi_I)
&={e^2\over 2}
\sum_A\biggl(\sum_I Q^I_A|\phi_I|^2-r_A\biggr)^2+
\sum_{A,B=1}^2\sum_I Q^I_AQ^I_B|\phi_I|^2\s_A\bar\s_B+\sum_I|J_I|^2
\cr
&={e_1^2\over 2}
\biggl(|z_1|^2+|z_2|^2+2|z_3|^2+2|z_4|^2+2|z_5|^2-8|p|^2-r_1\biggr)^2\cr
&+{e_2^2\over 2}\biggl(|z_1|^2+|z_2|^2-2|x|^2-r_2\biggr)^2\cr
&+|\s_1+\s_2|^2\biggl(|z_1|^2+|z_2|^2\biggr)+
|\s_1|^2\biggl(64|p|^2+4|z_3|^2+4|z_4|^2+4|z_5|^2\biggr)\cr
&+4|\s_2|^2|x|^2
+|G|^2+\sum_i|J_i|^2+|J_x|^2}}
The first two terms on the right hand side come {}from integrating out
the auxiliary fields $D_A$.

The theory can be studied semiclassically at large $|r_A|$; for $r_A>>0$
we find the Calabi-Yau phase in which the linear sigma model describes
string propagation on the hypersurface of degree 8 \defGfull\ in ${\bf
WP}^{4}_{11222}$.  For generic values of $r_A$ and the parameters
defining the polynomials $J_I$, the model is nonsingular: there is a
nonvanishing potential as the fields become large in any direction in
field space.\foot{As discussed in \SilvWit, it is not true that the
potential {\it grows} in every direction in field space, as setting
$\phi_I=0$ leaves a constant action ${{\rm (Area~of~worldsheet)}\times
e^2\over 2}\sum_A(r_A^2+ ({\theta_A\over{2\pi}}^2))$ \SilvWit.  This
does not lead to a divergence in the path integral for the physical
limit in which the area of the worldsheet goes to infinity.}

At complex codimension one in the linear sigma model
moduli space one finds a singular locus for which a direction
in field space exists with vanishing potential for large field 
strength.  For example, when $t_2\to 0$, $\s_2$ can become
arbitrarily large with no cost in potential as long as
$z_1=z_2=x=0$.  It is another singularity which will be of interest
to us: the one which appears where $t_1=t_2$.  There 
$\s_1$ and $\s_2$ can become large with no cost in potential
as long as $\s_1+\s_2=0=z_3=z_4=z_5=p=x$.  

One quantity of physical interest in the low-energy effective $N=1$
supergravity theory is the set of Yukawa couplings of the generations
and antigenerations charged under the unbroken part of the $SO(32)$
gauge group.  In the present case these are the couplings ${\bf 26}_{\pm
1}{\bf 1}_{\mp 2}{\bf 26}_{\pm 1}$ where we have indicated the gauge
charges under the unbroken $SO(26)\times U(1)$.  In our example there
are 2 generations (related on the (2,2) locus by left-moving
supersymmetries to the Kahler moduli $r_1$ and $r_2$) and 86
antigenerations. As discussed in \SilvWit, the linear sigma model vertex
operators for the generations are (linear combinations of) $\s_1$ and
$\s_2$.  The simple Hamiltonian computation of \SilvWit\ reveals that
the cubic coupling of $\s_1-\s_2$ has a simple pole in $t_1-t_2$ as
$t_1-t_2\to 0$ as in \yukpole.  For (0,2) models, singularities which
occur in the large radius phase generically involve the vector bundle V
degenerating while the manifold remains smooth.  Singularities such as
the present one which occur in the (worldsheet) quantum regime in (0,2)
models are not accessible by a (0,2) generalization of mirror symmetry
but can be studied fruitfully in the linear sigma model as just
indicated.  On the (2,2) locus, one finds a logarithmic singularity in
the metric on the moduli space; the behavior of the metric is not known
for (0,2) models.

\subsec{The Theory on the Generic Fiber}

The manifold \defGfull\ is a K3 fibration, as can be
seen {}from the field redefinition
\eqn\fibcoor{z_1=\lambda z_2,~~
\lambda_-^1=\lambda\lambda_-^2,~~ y\equiv xz_2^2}
upon which the defining equation becomes
\eqn\fibdef{G_f(y,z_3,z_4,z_5)={{(1+\lambda^8)y^4}\over 8}+
{z_3^4\over 4}+{z_4^4\over 4}+{z_5^4\over 4}
+\dots=0.}
This is the defining equation for the quartic hypersurface in
${\bf CP}^3$, one algebraic realization of K3.

In the heterotic context the fiber theory consists of the
pair (K3, $V_f$), where $V_f$ is the vector bundle to which
the left movers in the fiber theory couple.  Therefore
in addition to deducing the fiber manifold \fibdef, we
must also extract the fiber theory's vector bundle polynomials
(i.e. the K3 theory's version of the $J_I$ in the Calabi-Yau theory),
which we will denote $F_1,\dots,F_4$.  Let us work this
out starting {}from a general vector bundle in the Calabi-Yau
theory:
\eqn\Jone{J_1=p[z_1^7x^4+G_1]}
\eqn\Jtwo{J_2=p[z_2^7x^4+G_2]}
\eqn\Jthr{J_3=p[z_3^3+G_3]}
\eqn\Jfou{J_4=p[z_4^3+G_4]}
\eqn\Jfiv{J_5=p[z_5^3+G_5]}
\eqn\Jx{J_x={p\over 2}[(z_1^8+z_2^8)x^3+G_x]}
We require here that \EcondI\EcondII\ be satisfied.  This requires in 
particular that 
\eqn\EcondIIagain{2xJ_x=z_1J_1+z_2J_2.}
Using this, \fibcoor, and \sup, the worldsheet superpotential terms
involving $J_1$, $J_2$, and $J_x$ are
\eqn\supI{{\cal L}_{sup}=\int d\theta\biggl(\bigl(\Lambda_-^2+
{{\Lambda_-^xz_2}\over{2x}}\bigr)
\bigl(\lambda J_1 + J_2\bigr)\biggr)+\dots}
If we define 
\eqn\leftferm{\chi_-\equiv{\biggl({\Lambda_-^2
+{{\Lambda_-^xz_2}\over{2x}}} \biggr)z_2x}}
and
\eqn\Fone{F_1\equiv {{\lambda J_1+J_2}\over {z_2x}}\biggl|_{z_1=\lambda
z_2, 
y=xz_2^2}}
\eqn\Ftwo{F_2\equiv J_3(y, z_3,z_4,z_5)}
\eqn\Fthr{F_3\equiv J_4(y, z_3,z_4,z_5)}
\eqn\Ffou{F_4\equiv J_5(y, z_3,z_4,z_5)}
then the superpotential becomes
\eqn\supall{{\cal L}_{sup}=\int d\theta\biggl(\Lambda_-^pG_f+
\chi_-F_1+\Lambda_3F_2
+\Lambda_4F_3+\Lambda_5F_4\biggr).}
Note that the definition \Fone\ makes sense since for gauge invariance
each term in $G_1$ and $G_2$ must contain at least one factor
of $z_1$ or $z_2$ and at least one factor of x.  
In particular, \Fone-\Ffou\ reduce to the tangent
bundle of the K3 fiber when $G_i=0$ in the Calabi-Yau theory 
\Jone-\Jx. 

This superpotential determines the 
manifold and rank 2 bundle for the geometrical phase of large 
$r_1$. (Here $r_2$ is of course taken to be large, as we are considering
an adiabatic limit in which we can study the theory
on the generic fiber over the ${\bf P^1}$ whose
size is given by $r_2$ \bospot.)  For $t_1-t_2$ small, which will be near
the singularity of interest, this linear sigma model is strongly
coupled and does not directly give a description in terms of classical
geometry.  Because we are on K3, which generically has no
worldsheet instantons, this is only an artifact of
the description, and in the infrared the model must reduce to
some conformal field theory on K3 with
a rank 2 vector bundle.  

We would like to set up the model
so that the theory on the generic
fiber has a small instanton singularity when $t_1-t_2\to 0$, 
so that we can use the results of \sminst.
This will involve turning on the (0,2) moduli in the $J_I$, 
and hence (0,4) moduli in the fiber theory, since
we are aiming for a singularity of the vector bundle on
the fiber which is not accompanied by a singularity of the       
K3 manifold itself.    

First consider the (4,4) locus, which we can study
explicitly using mirror symmetry to map us to a model in
which the classical geometry is evident, since K3 is self-mirror.
One example of a description which we obtain in this way is the
following (using the procedure discussed in \CDGP\ for
the quintic threefold in ${\bf CP}^4$, translated one dimension
down to the quartic twofold in ${\bf CP}^3$).  
To obtain a mirror description we mod out
by all phase symmetries preserving the holomorphic 2-form \GP.
Set $(y,z_3,z_4,z_5)\equiv 
(\eta(\lambda)x_1,\dots,x_4)$, where
$\eta(\lambda)=(8/(1+\lambda^8))^{1\over 4}$.  A basis for these
phase symmetries is $g_1:(x_1,\dots, x_4)\to 
(x_1, \alpha x_2, x_3,\alpha^3 x_4)$ and 
$g_2:(x_1,\dots,x_4)\to (x_1,x_2,\alpha x_3, \alpha^3 x_4)$ where
$\alpha=exp({{2\pi i}\over 4})$.
This restricts us to one deformation theoretic modulus $\rho-1$, which
corresponds to the parameter $t_1-t_2$ in the original linear
sigma model description:
\eqn\mireq{G_{f,orb}={x_1^4\over 4}+{x_2^4\over 4}+
{x_3^4\over 4}+{x_4^4\over 4}-\rho x_1x_2x_3x_4=0.}
We now make the following field redefinition
\eqn\xtoy{(x_1,\dots, x_4)\equiv (y_1y_3^{1\over 4},
y_2^{3\over 4}y_4^{1\over 4},y_3^{3\over 4}y_2^{1\over 4},y_4^{3\over
4})} 
which respects the phase symmetries.  Then in terms of 
$(y_1,\dots,y_4)$, the defining equation becomes  
\eqn\mirfibdef{\tilde G_f(y_1,\dots,y_4)={y_1^4y_3\over 4}+
{y_2^3y_4\over 4}+{y_3^3y_2\over 4}
+{y_4^3\over 4}-\rho y_1y_2y_3y_4=0.}
This describes the same K3 as a hypersurface of degree 27 in
${\bf WP}^3_{5,6,7,9}$.  

Consider the following 2-parameter family of vector bundles:
\eqn\mirvecI{\tilde F_1=y_1^3y_3-\rho y_2y_3y_4}
\eqn\mirvecII{\tilde F_2={3\over 4}y_2^2y_4+{y_3^3\over 4}
-\rho y_1y_3y_4+{\delta\over 4}y_3^3}
\eqn\mirvecIII{\tilde F_3={3\over 4}y_3^2y_2+{y_1^4\over 4}
-\rho y_1y_2y_4-{\delta\over 4}y_2y_3^2}
\eqn\mirvecIV{\tilde F_4={y_2^3\over 4}+{3\over 4}y_4^2
-\rho y_1y_2y_3}
Here $\delta$ is a (0,4) deformation which preserves the fact
that the vector bundle satisfies the conditions analogous to
\EcondI\EcondII\ in this case.  The singularity on the
(4,4) locus occurs at $(\rho,\delta)=(1,0)$.  Starting {}from this
locus, turning
on $\rho-1$ alone removes the singularity.  A little algebra
shows that turning on $\delta$ alone also removes the
singularity. Since (as is
clear {}from the original linear sigma model for the fiber) the
singularity is at codimension one in the linear sigma model moduli
space, this 
means that some combination of $\rho-1$ and $\delta$ preserves
the singularity.  Since turning $\rho-1$ on removes the singularity
{}from the manifold, the resulting fiber theory has a singularity
in the vector bundle $\tilde V_f$ and not the manifold \mirfibdef.

It is easy to check that each of the 45 hypermultiplet moduli of
the SU(2) gauge bundle on K3 has a complex deformation-theoretic
representative in the original linear sigma model.  This means
in particular that the deformation $\delta$ is accessible in
the linear sigma model.  
Thus by turning on (0,2) moduli in the Calabi-Yau theory
before taking the limit $t_1-t_2\to 0$, we can obtain a singularity
which on the generic fiber is a singularity of the bundle but
not the manifold.  Singularities of the gauge bundle on
K3 occur when one or more instantons shrink to zero size.  
Since the singularity is codimension one in the linear sigma model,
we expect there to be a single small instanton in the generic fiber
for $t_1-t_2\to 0$.\foot{This is true unless the linear sigma model
has ``blown up'' the singularity.  But with a rank 2 bundle there
are physical singularities at codimension one in the instanton 
moduli space.  So if the codimension one singularity we see in
the linear sigma model has been ``blown up'', then there are
others which have been ``blown down'' with respect to the
physical moduli space.  We will here assume that in the rank
2 case the linear sigma model gives an accurate description 
of the singularities.}

\subsec{Computation of the Splitting Type of V}

As discussed in the last section, we need to compute
the splitting type of V at the singularity in order to obtain
$H^0({\cal O}(-1)\otimes V)$ and $H^0({\cal O}(-1)\otimes V^*)$ there.
Because the left-moving fermions transform as sections
of the spinor bundle on the worldsheet and as sections
of V in spacetime,
this is the same as determining the zero modes of the
left-moving fermions of the linear sigma model 
in the background of one worldsheet instanton
on the curve.  
In particular, we can compute the precise splitting type of V by 
the methods used in \Katz\ for the quintic tangent bundle.
The ${\bf P^1}$ of interest is 
\eqn\inst{(z_1,\dots,z_5;x)=(\lambda z_2,z_2,a,b,c;0)}
Where $a$, $b$, and $c$ are constants satisfying
${a^4\over 4}+{b^4\over 4}+{c^4\over 4}=0$.
This satisfies \defGfull\ and gives the base of the fibration described
in the last section.  The structure of worldsheet instantons
in the linear sigma model was analyzed in \refs{\phases, \MP}.
The instanton \inst\ has 
\eqn\wsinst{\int v^{(2)}_{12} = 2\pi ~~;~~
\int v^{(1)}_{12}- v^{(2)}_{12} = 0.}
where $v_{12}^{(A)}$ is the field strength for the worldsheet gauge
group $U(1)_A$.


This means that the only fermions that can possibly have
zero modes are those charged under $U(1)_2$, since the
spinor bundle alone is ${\cal O}(-1)$ and $H^{0}({\cal O}(-1))=0$.
In particular, only $\lambda^1$, $\lambda^2$, and $\lambda^x$ are
charged under $U(1)_2$.  Since 
$J_1\bigl|_C=J_2\bigl|_C=0$,\foot{This holds even for a general
(0,2) deformation because for gauge invariance, each term
in $J_1$ and $J_2$ contains at least one power of $x$.}
$\lambda^1$ and $\lambda^2$ reduce at low energies in the linear
sigma model to a vector bundle transforming
as does the tangent bundle 
of ${\bf P}^1$ (tensored with the spinor bundle
${\cal O}(-1)$), and therefore lie in ${\cal O}(2)\otimes 
{\cal O}(-1)$ on C.
Similarly, since $\lambda^3$, $\lambda^4$, $\lambda^5$ transform
as sections of ${\cal O}(0)\times {\cal O}(-1)$ on C, and since
$J_3$, $J_4$, and $J_5$ are constant on the curve, we can solve
$\lambda^3J_3+\lambda^4J_4+\lambda^5J_5=0$ with
$\lambda^I\sim\lambda^I+Q_A^I\Phi^I\eta$, and there is therefore 
a term ${\cal O}(0)$ in $V\bigl|_C$.  
Since $c_1(V)=0$, and $V$ is rank 3, we immediately have that 
$V\bigl|_C={\cal O}(2)+{\cal O}(0)+{\cal O}(-2)$ as promised in 
\Vsplit.  (This is the splitting type one would have for
the tangent bundle, since there is a curve of ${\bf P^1}$'s in \wpsing.)
Thus we are left with SU(2) with four doublets as
the nonperturbative enhancement of the spectrum
for the $N=1$ theory at the ``conifold'', as anticipated.

\medskip
\noindent {\bf Acknowledgements:}

We would like to thank S. Katz, D. Morrison, and E. Witten
for very useful discussions.  
The research of S.K. was supported in part by the Harvard Society
of Fellows.  The research of N.S. was supported in part by DOE grant
\#DE-FG02-96ER40559.

\listrefs
\end